\documentclass{article}

\usepackage{arxiv}

\usepackage[utf8]{inputenc} 
\usepackage[T1]{fontenc}    
\usepackage{hyperref}       
\usepackage{url}            
\usepackage{booktabs}       
\usepackage{amsfonts}       
\usepackage{nicefrac}       
\usepackage{microtype}      
\usepackage{lipsum}

\usepackage[pdftex]{graphicx}
\graphicspath{{./figures/}}
\usepackage{amsmath}

\usepackage{amsthm}

\usepackage{algorithm}
\usepackage{algpseudocode}

\newcommand\concat{\mathbin{+\mkern-10mu+}}

\title{Group Testing for Efficiently Sampling Hypergraphs When Tests Have Variable Costs\\}

\author{
  Laurence A.~Clarfeld\\
  Department of Computer Science\\
  University of Vermont\\
  Burlington, VT 05405 \\
  \texttt{Laurence.Clarfeld@uvm.edu} \\
   \And
 Margaret J.~Eppstein \\
  Department of Computer Science\\
  University of Vermont\\
  Burlington, VT 05405 \\
  \texttt{Maggie.Eppstein@uvm.edu} \\
}

\begin{document}
\maketitle

\begin{abstract}
In the group-testing literature, efficient algorithms have been developed to minimize the number of tests required to identify all minimal ``defective'' sub-groups embedded within a larger group, using deterministic group splitting with a generalized binary search. In a separate literature, researchers have used a stochastic group splitting approach to efficiently sample from the intractable number of minimal defective sets of outages in electrical power systems that trigger large cascading failures, a problem in which positive tests can be much more computationally costly than negative tests. In this work, we generate test problems with variable numbers of defective sets and a tunable positive:negative test cost ratio to compare the efficiency of deterministic and stochastic adaptive group splitting algorithms for identifying defective edges in hypergraphs.  For both algorithms, we show that the optimal initial group size is a function of both the prevalence of defective sets and the positive:negative test cost ratio. We find that deterministic splitting requires fewer total tests but stochastic splitting requires fewer positive tests, such that the relative efficiency of these two approaches depends on the positive:negative test cost ratio. We discuss some real-world applications where each of these algorithms is expected to outperform the other.
\end{abstract}


\section{Introduction}
\label{sec:introduction}

The field of group testing is thought to have originated from a single report by Dorfman in which he proposed a novel method for efficiently screening soldiers for syphilis during World War II \cite{dorfman1943detection}. Dorfman suggested mixing together blood samples from multiple individuals so it would require just a single chemical test to determine if the pooled blood sample contained syphilitic antigen. If the test came back negative, it would indicate that none of the soldiers were infected, whereas a positive test result would require subsequent tests to determine which soldiers were infected. Although this initial proposal was never implemented, group testing has since been applied to solve combinatorial search problems in a variety of disciplines \cite{du2000combinatorial}.  Group testing strategies are currently being explored to minimize the number of tests needed estimate the prevalence of Covid-19 in large swaths of the population ({\em e.g.}, \cite{yelin2020evaluation,mentus2020analysis}).
Finding optimal group testing strategies that minimize the number of tests required to identify ``defective'' individuals (or items, or minimal sets of items) has been a central focus in the robust group-testing literature that has emerged. An implicit assumption of this work is that the cost of positive tests is the same as the cost of negative tests.  In this paper, we examine so-called ``adaptive'' group testing algorithms for finding defective edges in hypergraphs, where the results of previous tests are used to inform which tests to perform next.

A hypergraph is a generalization of an ordinary graph, in which individual edges (so-called ``hyperedges'') can connect an arbitrary number of nodes. Specifically, define a hypergraph $G(V,E)$ where each node in $V$ represents one of the $N$ individuals or items of interest and each hyperedge in $E$ connects a set of $k$ nodes, or $k$-set. We consider the most general case, where there exists an edge in $E$ for every possible subset of $V$.  When using group-testing to identify defective edges in a hypergraph, the aim is to identify minimal defective $k$-sets in $E$, where ``minimal'' means that no smaller subset is defective. 

 Searching for defective edges in a graph $G$ with $|E|$ edges using group testing was first considered by Aigner, who conjectured that no more than $\lceil\log_2|E|\rceil+c$ tests (for some constant $c$) were required to find a single defective edge in $G$ \cite{aigner1988combinatorial}. This was later proven by Damaschke \cite{damaschke1994tight} and generalized to hypergraphs by Triesch \cite{triesch1996group}. The case of finding all $d>1$ defective edges in a graph, where $d$ is known, was first addressed in \cite{johann2002group}. For the case when $d$ is unknown, adaptive methods for finding all defective edges were proposed for graphs in \cite{hwang2005competitive} and extended to hypergraphs in \cite{chen2007competitive,chen2011revised}. All of the aforementioned algorithms use deterministic splitting approaches. 

One important real-world problem that can be framed as searching for defective hyperedges in a hypergraph is the identification of minimal sets of $k$ outages that trigger cascading failures in power systems, as a means of estimating overall risk of cascading failure. This real-world problem has several characteristics that have not traditionally been considered in the standard group-testing literature: (i) The number of minimal defective edges $d$ is so large that it is computationally intractable to identify them all, so sampling is required;  (ii) $G$ is non-uniform (\textit{i.e.}, $k$ is variable, in the $k$-sets defined by the hyperedges in $E$); (iii) the size of the defective sets sought is lower-bounded by $k_{min}=2$, since power systems are operated such that no single outage ($k=1$) will result in a cascading failure; (iv) the size of the defective sets sought is often upper bounded by $k_{max}$, due to computational limitations;  (v) tests can produce false negatives, wherein a set tests as non-defective even though it contains a defective subset, (\textit{e.g.}, this can occur when a power grid is fragmented into disconnected but independently functioning ``islands'' \cite{aghamohammadi2012intentional}); and, most importantly, (vi) positive and negative tests may have very different costs.

\begin{figure}[!t]
\centering
\includegraphics[width=3.5in]{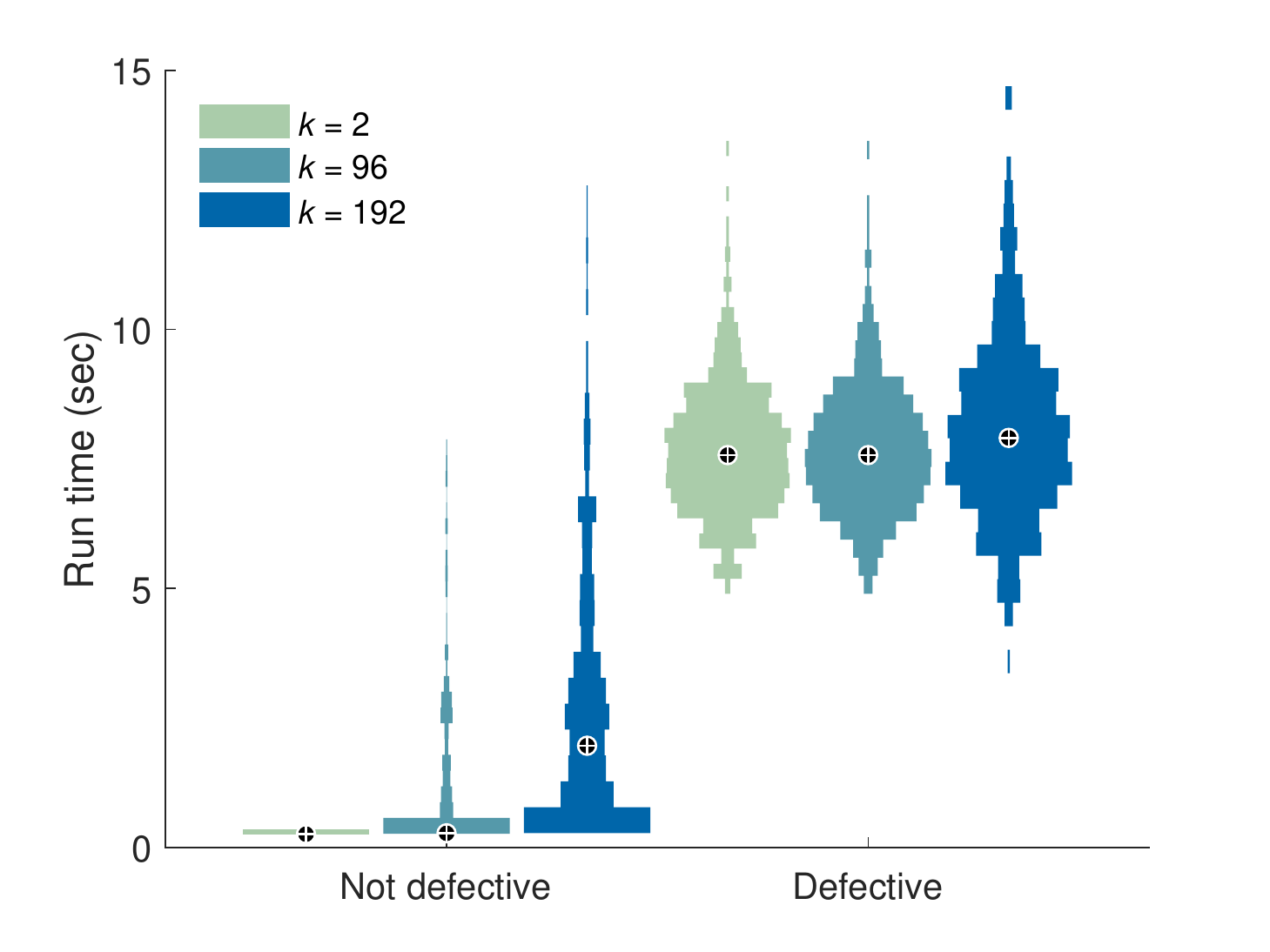}
\caption{Run times required using DCSIMSEP to test 500 random non-defective sets and 500 random defective sets, for each of set sizes $k \in \{2, 96, 192\}$,  on a synthetic model of the Western U.S. power grid. To generate this data, a set was considered defective if there was at least 5\% of the load shed in DCSIMSEP. For clarity, medians are marked with crosshairs and each distribution has been independently normalized to the same maximum width. See Sec. \ref{sec:TEST_PROBLEM} for a description of the simulator and the test case.}
\label{fig:pos_vs_neg}
\end{figure}

The reason that test costs are so variable in the power systems problem is that simulating each step of a cascade accrues additional computational costs, and positive tests have larger cascades than negative tests, by definition. We illustrate an example of this difference in the costs of positive and negative tests using DCSIMSEP ~\cite{Hines2016cascadingChapter} (Fig. \ref{fig:pos_vs_neg}), an open-source DC simulator that has been used in studying cascading power outages \cite{eppstein2012random,rezaei2014estimatingC, rezaei2014estimating, rezaei2015rapid, clarfeld2018assessing, clarfeld2019risk}. We note that a variety of types of DC and AC power flow solvers and cascade models could theoretically be used to simulate cascading failures in power grids, each with their own sets of advantages and disadvantages \cite{haes2019survey}, \cite{ju2018modeling}, \cite{yang2017small}, \cite{guo2017critical}, \cite{nagarajan2017resilient},\cite{cetinay2017comparing}, \cite{bent2014transmission}. If one were to use a more sophisticated AC simulator, the relative difference between the computational cost of positive and negative tests would be expected to be even greater, because each step of an AC simulator will be more computationally costly than in a DC simulator \cite{ju2018modeling}.

In the power systems literature, one method that has been proposed to tackle the problem of efficiently finding minimal defective $k$-sets is the Random Chemistry (RC) algorithm \cite{eppstein2012random, rezaei2014estimatingC, rezaei2014estimating, rezaei2015rapid, clarfeld2018assessing, clarfeld2019risk}. RC is a stochastic adaptive group testing approach. To our knowledge, deterministic group-testing approaches have not previously been applied to the power systems problem or other problems where the cost of positive and negative tests is unequal. 

Although this work was originally motivated by the power systems problem described above, there are other potential applications that share many of these characteristics, including variable positive:negative test costs (see Sec. \ref{sec:discussion}).  Thus, the aim of this study is to explore the general question regarding the relative computational efficiency of deterministic and stochastic adaptive group-testing algorithms for sampling minimal defective sets, in problems where there are large and unknown minimal defective $k$-sets with variable $k$, the potential for false negatives exists, and in which the relative computational costs of positive and negative tests may differ. 

To that end, we create test problems with different numbers of minimal defective hyperedges, low (but non-zero) frequencies of false negatives, and tunable positive:negative test costs. We compare the RC stochastic adaptive group-testing algorithm to a deterministic adaptive group-testing algorithm we refer to as SIGHT (Sampling Inspired by Group Hyperedge Testing). SIGHT is a minor adaptation of traditional group testing strategies that have been designed to minimize the number of required tests to find all minimal defective sets \cite{triesch1996group,hwang2005competitive,chen2007competitive}, with small modifications to turn it into a sampling algorithm. Specifically, unlike \cite{triesch1996group,hwang2005competitive,chen2007competitive}, SIGHT (i) searches for only one of an unknown number $d$ of minimal defective $k$-sets, for $k_{min} \le k \le k_{max}$, and (ii) is tolerant of false negatives ({\em i.e.}, where a $k$-set that tests as non-defective actually contains at least one minimal defective $k$-set). 

This paper is organized as follows: The SIGHT algorithm is described in Section \ref{sec:SIGHT}, the RC algorithm in Section \ref{sec:RC}, and the test problem generator is described in Section \ref{sec:TEST_PROBLEM}. Experiments to compare the relative performance of the two algorithms for finding minimal defective $k$-sets are described in Section \ref{sec:experiments}, with results presented in Section \ref{sec:results}. We discuss the implications of our findings for some important applications in Section \ref{sec:discussion} and summarize our conclusions in Section \ref{sec:conclusion}. Although this study was originally motivated by the power systems problem described above, the implications are relevant to any hypergraph sampling problem. 

\section{Methods}

\subsection{SIGHT} \label{sec:SIGHT}
\subsubsection{Group Testing Inspiration}
In the simple case of a single minimal defective edge in a graph, Triesch proposed a group testing halving procedure \cite{triesch1996group} that was later adapted by Johann \cite{johann2002group} to find all $d$ defective edges (2-sets) in a simple graph in at most $d(\lceil \log_2\frac{m}{d}\rceil+7)$ tests, where $d$ is known. The algorithm presented by Johann for finding a single minimal defective edge in a subgraph $G' \subset G$ has subsequently been referred to as the TJ Procedure \cite{hwang2005competitive,chen2007competitive} and is the primary inspiration for the $BinSearchSIGHT$ subroutine presented in Sec. \ref{sec:SIGHT_alg}. The TJ procedure was first used to find all $d$ defective edges, when $d$ is unknown, in simple graphs \cite{hwang2005competitive}, and then extended to hypergraphs \cite{chen2007competitive, chen2011revised}.  SIGHT is a minor adaptation of the TJ procedure, modified in two ways: (i) it is designed to randomly sample a single defective $k$-set for $k_{min} \leq k \leq k_{max}$ (as opposed to assuming all defective sets of all sizes will be found), aborting if the minimal defective set has $k>k_{max}$; and (ii) it is designed to be robust to false negative tests.

\subsubsection{SIGHT Algorithm} \label{sec:SIGHT_alg} 
SIGHT takes as input the universal set $V$ of nodes,  an initial subset size $a_0$, and the bounds $k_{min}$ and $k_{max}$ on the size of the defective $k$-sets one is searching for. It returns either a minimal defective $k$-set (for $k_{min} \leq k \leq k_{max}$) or the empty set (if the algorithm aborts). The only control parameter required by SIGHT is $a_0$, which establishes the size of the initial subset to be tested, where $k_{max} \leq a_0 < N$. The algorithm is illustrated by the Psuedocode in Algorithm \ref{alg:SIGHT} and is described below. 

\begin{algorithm}
\caption{SIGHT$(V,a_0,k_{min},k_{max})$} \label{alg:SIGHT}
\begin{algorithmic}[1]
    \State $S\gets SAMPLE(V,a_0)$
    \State $D\gets \emptyset$
    \If {$\neg \, isDefective(S)$}
        \State \textbf{return} $\emptyset$
    \EndIf
    
    \While {$|D| < k_{max}$}
        \State $m \gets BinSearchSIGHT(S,D)$
        \State $D\gets D \concat S[m]$
        \If {$|D| \geq k_{min} \; \land isDefective(D)$ } \footnotemark
            \State $D \gets bottomUpSIGHT(D,k_{min},k_{max})$
            \State \textbf{return} $D$
        \EndIf
        
        \State $S\gets S[{\;:\,}m-1]$ 
    \EndWhile
    \State \textbf{return} $\emptyset$
\end{algorithmic}
\end{algorithm}
\footnotetext{The use of short-circuiting logic in the conditional statement prevents unnecessary tests.}

The first step of the SIGHT algorithm  (Alg. \ref{alg:SIGHT}: line 1) calls the subroutine $SAMPLE(V,a_0)$ (pseudocode not shown), which initializes a list $S$ to a uniform random sample, in random order, of $a_0$ unique elements from $V$. In the following description, the notation $S[i]$ is used to refer to the $i^{\mathrm{th}}$ element of $S$; $S[i]$ is considered to be left of $S[j]$ for all $i<j$; and $S[{\;:\,}i]$ refers to the first $i$ elements of $S$.

A list $D$ is initialized to the empty list (Alg. \ref{alg:SIGHT}: line 2). This list $D$ will be used to accumulate known nodes from defective $k$-sets.
In Alg. \ref{alg:SIGHT}: line 3, the subroutine $isDefective(S)$ is called to test whether the list $S$ is a defective set (although not necessarily minimal). The algorithm $isDefective(S)$ is specific to the application problem, so is not shown here. It is \textit{expected} to return TRUE if there is a minimal defective $k$-set in $S$ and FALSE otherwise; however, if false negatives can occur then $isDefective(S)$ may sometimes return FALSE even when $S$ contains a minimal defective set. If $isDefective(S)$ returns FALSE, the algorithm aborts and returns the empty set (Alg. \ref{alg:SIGHT}: line 4). Otherwise, the algorithm calls the subroutine $BinSearchSIGHT(S,D)$ (Alg. \ref{alg:SIGHT}: line 7).

The subroutine $BinSearchSIGHT(S,D)$ (Algorithm \ref{alg:bin}), following \cite{johann2002group,chen2007competitive,hwang2005competitive},  uses a deterministic binary search to find the index of the leftmost element in $S$ that is the rightmost element of a defective $k$-set in $D \concat S$, where $\concat$ denotes list concatenation. The subroutine is implemented like a classic binary search, in that it first assigns left and right pointers $l$ and $r$ to the leftmost and rightmost elements of $S$, respectively (Alg. \ref{alg:bin}: lines 1-2), and determines a test index, $i$, half way between $l$ and $r$. But unlike a standard binary search that assumes that the values of the elements of $S$ are in sorted order and performs a test on element $S[i]$, here only the ordering of the indices matters (the actual elements in $S$ are deliberately in random order to prevent the algorithm from biasing towards sets that include elements with low values)  and the test (Alg. \ref{alg:bin}: line 5) includes all elements of $S$ with indices $\le i$ as well as all elements from the growing list $D$, to see whether $D \concat S[{\;:\,}r-i]$, is defective. If so, $r$ is reduced (Alg. \ref{alg:bin}: line 6); if not, $l$ is increased (Alg. \ref{alg:bin}: line 8). The process repeats until $l$ and $r$ converge to some index, which is returned. 

\begin{algorithm}
\caption{BinSearchSIGHT$(S,D)$}  \label{alg:bin}
\begin{algorithmic}[1]
    \State $l \gets 1$
    \State $r \gets |S|$
    \While {$l<r$}
        \State $i \gets \lceil \frac{r-l}{2}\rceil$
        \If {$isDefective(D \concat S[{\;:\,}r-i])$}
            \State $r \gets r-i$
        \Else
            \State $l \gets r - i + 1$
        \EndIf
    \EndWhile
    \State \textbf{return} r
\end{algorithmic}
\end{algorithm}

After each call to the $BinSearchSIGHT$ subroutine, SIGHT concatenates the found element $S[m]$ to $D$ (Alg. \ref{alg:SIGHT}: line 8). If the accumulated set $D$ is found to be defective and is at least size $k_{min}$ (Alg. \ref{alg:SIGHT}: line 9), then the algorithm has successfully found a list $D$ that contains a minimal defective $k$-set. However, in applications where false negatives can occur, $D$ may be non-minimal. To make the algorithm more robust to false negatives, it then calls the subroutine $bottomUpSIGHT(D,k_{min},k_{max})$ (Alg. \ref{alg:SIGHT}: line 10; pseudocode for $bottomUpSIGHT$ not shown), which tests any subsets of $D$ of size $k_{min} \le k \leq k_{max}$ that have not already been tested, to ensure that the defective subset returned is minimal  (Alg. \ref{alg:SIGHT}: line 11). If $D$ is not defective, then $S[m]$ and all elements to its right are removed from $S$ (Alg. \ref{alg:SIGHT}: line 13) and the process is repeated until either $D$ tests as defective, or $|D|$ exceeds $k_{max}$ (Alg. \ref{alg:SIGHT}: line 6), in which case the defective set being found is too large and the algorithm aborts (Alg. \ref{alg:SIGHT}: line 15). Note that Alg. \ref{alg:SIGHT}: lines 6, 9-12 comprise the minor differences between SIGHT and the previous approach to group testing on hypergraphs given in ~\cite{chen2007competitive}, on which SIGHT is based. Open source Matlab code for SIGHT is posted online \cite{SIGHTcode}.


\subsubsection{Computational Complexity of SIGHT}

The time complexity of $isDefective(S)$ is application-dependent, so computational complexity is here defined as a function of the number of tests required (\textit{i.e.}, calls to $isDefective(S)$). Each call to $BinSearchSIGHT$ takes no more than $\lceil \log_2(a_0) \rceil$ tests to find an element of the defective set, and this operation is performed at most $k_{max}$ times before either a defective set is found or the algorithm aborts. Additionally, all subsets of $D$ larger than $k_{min}$ must be tested to ensure the defective set returned is minimal. The resulting upper-bound on the number of required tests by SIGHT is thus:

\begin{align} \label{t_max_SIGHT}
    max(\#Tests|SIGHT) =  k_{max}\lceil\log_2(a_0)\rceil+\sum_{j=k_{min}}^{k_{max}} {k_{max} \choose j}+1
\end{align}

In practice, this worst case is rarely realized. Since $a_0 < N$, each run of SIGHT requires $O(log N)$ tests.

However, not all tests are necessarily equal. In some applications, such as the power systems application described earlier, positive tests require much more computation time than negative tests.  The required number of positive tests per each run of SIGHT is upper-bounded by: 

\begin{align}\label{eq:maxposSIGHT}
    max(\#Positive Tests | SIGHT) = k_{max} \lceil \log_2 (a_0)\rceil
\end{align}
Note that this is explicitly a function of $k_{max}$.

\subsection{Random Chemistry} \label{sec:RC}

\subsubsection{Inspiration for RC}
The basic idea for the RC algorithm was originally proposed by Kauffman \cite{kauffman1996home} as a hypothetical method for identifying minimal auto-catalytic sets of interacting molecules from within a very large set of molecules. He suggested testing random half-sets until the products of auto-catalysis were detected in one subset, and repeating this halving process until a minimal auto-catalytic set was discovered (hence the moniker ``Random Chemistry''). Although not previously presented as such, this is an adaptive stochastic group testing method.

Inspired by the RC idea, an RC algorithm was implemented for finding genetic interactions between single nucleotide polymorphims that predispose for disease \cite{eppstein2007genomic}. Later, a version of RC was implemented for finding a small set of transmission line outages in electric power networks that trigger cascading power failure, requiring only $O(log N)$ tests per successful run~\cite{eppstein2012random}. 

\subsubsection{RC Algorithm}
As in SIGHT, RC takes as input the set of nodes $V$ and returns either a single minimal defective $k$-set (for $k_{min} \leq k \leq k_{max}$), or the empty set (if the algorithm aborts). The bounds on $k$ are specified for RC exactly as they are in SIGHT. Pseudocode for RC, as implemented in this study, is provided in Algorithm \ref{alg:RC}. 

\begin{algorithm}
\caption{RC$(V,A,k_{min},k_{max},t_{max})$} \label{alg:RC}
    \begin{algorithmic}[1]
    \State $a_0\gets A(0)$
    \State $S\gets SAMPLE(V,a_0)$
    \If {$\neg \, isDefective(S)$}
        \State \textbf{return} $\emptyset$
    \EndIf
    
    \For {$i \gets 1, |A|$}
        \State $a_{i} \gets A(i)$
        \State $t = 0$
        \State $flag \gets false$
        \While {$t < t_{max} \land \neg flag$}
            \State $t \gets t + 1$
            \State $S_{new} \gets SAMPLE(S,a_{i})$
            \If {$isDefective(S_{new})$}
                \State $S \gets S_{new}$
                \State $flag \gets true$
            \EndIf
        \EndWhile
        \If {$\neg flag$}
            \State \textbf{return} $\emptyset$
        \EndIf
    \EndFor
    \State \textbf{return} $bottomUpRC(S,k_{min},k_{max})$
    \end{algorithmic}
\end{algorithm}
~\\

Like SIGHT, RC begins by drawing a random sample $S$ of elements from $V$ such that $|S|=a_0$ (Alg. \ref{alg:RC}: lines 1-2) and aborts, returning the empty set, if $S$ does not contain a defective set (Alg. \ref{alg:RC}: lines 3-5). If an initial defective set $S$ of size $a_0$ is found, subset reduction proceeds stochastically according to the set size reduction scheme $A$ (loop starting at Alg. \ref{alg:RC}: line 6).

The sampling loop in RC (Alg. \ref{alg:RC}: lines 10-17) stochastically attempts to find a defective subset $S_{new}$ of size $a_{i}$, from set $S$ of size $a_{i-1}$. If no such subset is found after $t_{max}$ attempts, the algorithm aborts and returns the empty set (Alg. \ref{alg:RC}: lines 18-20). When a subset of size $a_{final}$ is found that causes a cascade, a bottom-up search is conducted (Alg. \ref{alg:RC}: line 22, which calls subroutine $bottomUpRC(S,k_{min},k_{max})$, pseudocode not shown), testing all subsets of size $k$, for $k=k_{min},\dots ,k_{max}$ (in random order for each $k$), returning either the first defective $k$-set found or the empty set, if no minimal defective set of size $\leq k_{max}$ exists in $S$. The subroutine $bottomUpRC$ differs slightly from $bottomUpSIGHT$, in that $bottomUpRC$ must test all subsets of $S$ until a defective subset is found or $k>k_{max}$. In SIGHT, some subsets of $S$ have already been tested during the binary search, so $bottomUpSIGHT$ only needs to test the subsets of $S$ that have not been previously tested. Open source Matlab code for RC, as implemented in this study, is posted online \cite{RCcode}.

If there is exactly one minimal defective $k$-set, and the cost of each test is constant, then one can derive an optimal set reduction scheme $A$ for RC ~\cite{buzas2013optimized}. However, in many applications (such as the power systems application) it is not possible to analytically optimize $A$. In the work shown here, we use the subset reduction schedule of $a_{i}=a_{i-1}/c$, where $c=2$ for $a_{i-1}>20$ (binary splitting) and $c=1.5$ for $a_{i-1} \le 20$ (to increase success rate when groups get small), and specify $t_{max}=20$, as proposed in \cite{eppstein2012random, rezaei2014estimating, rezaei2015rapid, clarfeld2018assessing, clarfeld2019risk}. The initial set size $a_0$ is tuned as a control parameter.

\subsubsection{Computational Complexity of RC}
The RC algorithm requires a test of the initial subset, plus up to $t_{max}$ tests for each of the $|A|-1$ reduction steps, plus additional tests for the bottom-up search of the set of size $a_{final}$. Thus, the maximum potential number of tests required by an RC run is:

\begin{align} \label{RC_big_O}
max(\#tests|RC) = 1+(|A|-1)t_{max} + \sum_{k=k_{min}}^{k_{max}} {a_{final} \choose k}
\end{align}

In practice, (using the RC parameters specified in the experiments presented here) only a few tests ($\ll t_{max}$) are typically required during each stage of the subset reduction, and $2$-sets are found most frequently (Section \ref{sec:results}) even when $k_{max}>2$, so the average number of tests required is much lower than this (Section \ref{sec:results}). As long as $a_{i}=a_{i-1}/c$, for some constant(s) $c > 1$ (as implemented here), then $|A| \propto log(a_0)$, for some $a_0 < N$. Under these circumstances, each RC run requires $O(log N)$ tests.

In RC, each reduction step requires exactly one defective test, so the number of defective tests required by a successful RC run is constant. For a reduction scheme of length $|A|$ followed by a brute force search until one defective set is found, the maximum number of tests of defective sets required by an RC run is: 
\begin{align}\label{eq:maxposRC}
    max(\#defective Tests | RC) = {|A|+1} 
\end{align}

Assuming set sizes in $A$ are fractionally reduced, \ref{eq:maxposRC} reduces to:
\begin{align}
    max(\#defective Tests | RC) \le \lceil \log (a_0)\rceil + 1
\end{align}

Thus, the maximum number of defective tests is not a function of $k_{max}$, as it is in SIGHT (compare to Eq. (\ref{eq:maxposSIGHT})). Given that there is some unknown probability of aborting during any reduction step, the average number is less than this. 

\subsection{Tunable Test Problem Generator}\label{sec:TEST_PROBLEM}
We leverage DCSIMSEP as a convenient means of generating domain-independent test problems with large and variable (but unknown) numbers of minimal defective $k$-sets (for $k \ge 2$). 
In a power systems application, DCSIMSEP works as follows. Given an initial set of $k$ outages (e.g., transmission line failures), DCSIMSEP iteratively checks to see if some pre-defined threshold $T$ for system failure is exceeded.  For modeling cascading power failures, $T$ might be some percent of the load that is shed \cite{rezaei2014estimating} (as illustrated in Fig. \ref{fig:pos_vs_neg}) or some percent of components that become separated from the grid \cite{eppstein2012random}. If the threshold $T$ is exceeded, the simulation terminates. If not, then the simulator calculates how the power generation would change, how power flow would be redistributed, and whether this would cause additional components to fail. These iterations continue until either the threshold $T$ is exceeded (a positive test), the system achieves a new equilibrium (a negative test), or some maximum amount of time has elapsed (also a negative test).  DCSIMSEP has been shown to yield low, but non-zero, frequencies of false negatives \cite{rezaei2014estimating}.  

For the experiments presented here, we run DCSIMSEP on a large synthetic power system, with a geographical topography based on the footprint of the 11-state western Unites States transmission system, which is included in the Electric Grid Test Case Repository~\cite{birchfield2017grid}. The test case contains 10,000 buses (connection points, typically substations, through which generators provide power and loads draw power from the network) and $N=12,706$ branches (transmission lines and transformers). We limit our simulations to those in which $k$-sets comprise sets of extrinsically caused initiating branch outages, and we specify initial conditions that guarantee that the system is $N-1$ secure, using the approach described in \cite{rezaei2014estimating}; {\em i.e.} there are no defective 1-sets.   
 
For the purposes of using DCSIMSEP as a general test problem generator, rather than to model which $k$-sets cause cascading failures in power systems, we consider a set defective if DCSIMSEP simply exceeds a specified number of iterations ($T$). Due to the intractable number of unique $k$-sets, it is not computationally feasible to ascertain the exact numbers of minimal defective $k$-sets in these problems. However, because the number of minimal defective $k$-sets will decrease as $T$ increases, we are able to create test problems with varying numbers of minimal defective $k$-sets with $k \ge 2$.

Rather than using runtimes from DCSIMSEP as a measure of the variable computational costs of positive and negative tests, we assume that negative tests always require 1 unit of time to run and that all positive tests require some pre-specified constant units of time.
 
The resulting test problems and tunable P:N test cost ratios enable us to compare the relative efficiency of the SIGHT and RC sampling algorithms on arbitrary problems with different numbers of defective $k$-sets and different P:N test cost ratios.
 
\subsection{Experiments} \label{sec:experiments}

We report on results from 720,000 paired runs ({\em i.e.}, starting from the same random initial sets) of SIGHT and RC, with $k_{max} = 4$. Specifically, we report on 30,000 paired runs for each initial set size $a_0 \in \{16,48,80,112,144,176\}$ and each test problem threshold $T \in \{5,15,25,35\}$.  (We note that we also computed results for $k_{max} \in \{2,3\}$, but only present results for $k_{max}=4$, since the observed patterns were nearly identical for $k_{max} \in \{2,3\}$ and $k_{max}=4$ is the more general case.) 

We record the number of positive and negative tests required for each run, where a run comprises a single call to SIGHT (Alg. \ref{alg:SIGHT}) or RC (Alg. \ref{alg:RC}) on a given combination of $a_0$ and $T$ and either terminates successfully in a {\em find} (when a minimal defective $k$-set, with $k \in \{2,3,4\}$ is found) or is aborted. However, rather than reporting metrics {\em per run} of SIGHT or RC, we report metrics {\em per find}, as the latter amortizes in the cost of unsuccessful (aborted) runs by each sampling algorithm. 

Since the resulting distributions of the number of positive tests and negative tests required per find of a defective $k$-set ($k \in \{2,3,4\}$) were non-Gaussian, we used non-parametric Mann Whitney $U$ tests to assess whether median values were significantly different between SIGHT and RC. Based on the median number of positive and negative tests per find on each test problem, we compute the expected computational costs of each algorithm at various P:N test cost ratios. Specifically, we assume that the cost of each negative test is 1 unit of time and the cost of each positive test is one of $\{1, 10, 50, 100\}$.

\section{Results} \label{sec:results}

\begin{figure*}[!ht]
\centering
\includegraphics[width=\textwidth]{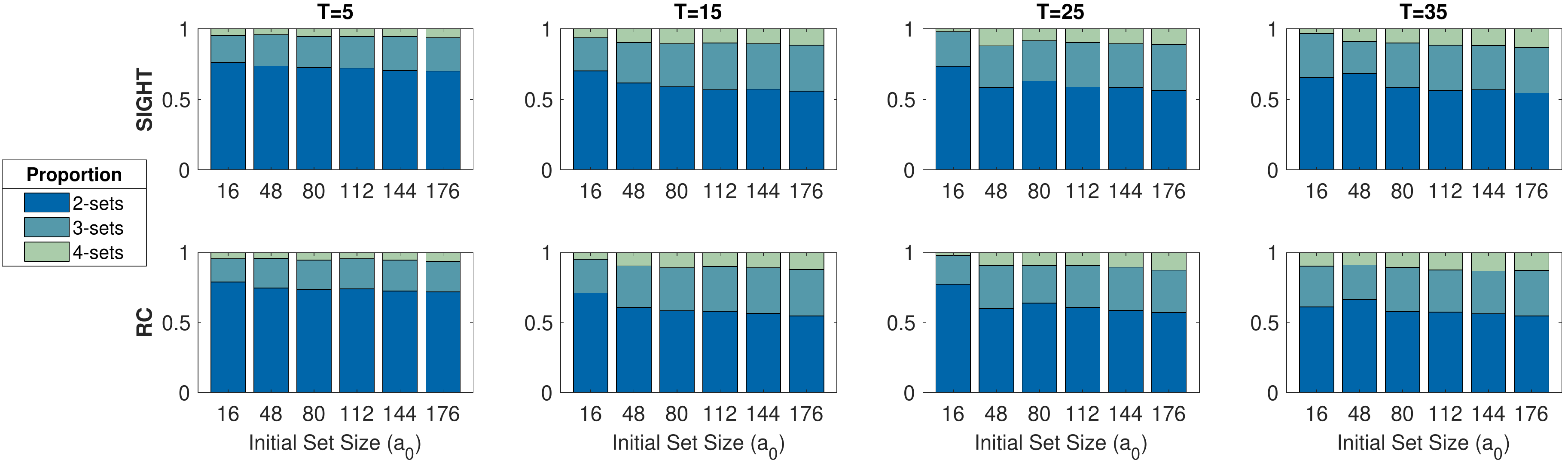}
\caption{The proportion of $k$-sets of size $k \in \{2,3,4\}$ found in 30,000 paired runs of SIGHT (top) and RC (bottom) with $k_{max}=4$, for each $a_0 \in \{16,48,80,112,144,176\}$ and each $T \in \{5,15,25,35\}$. }
\label{fig:defSetSize}
\end{figure*}

RC and SIGHT found similar proportions of minimal defective 2-sets, 3-sets, and 4-sets over all initial set sizes $a_0$ and all thresholds $T$ tested (Fig. \ref{fig:defSetSize}). The set size reduction approaches of both group-testing algorithms result in a bias toward finding smaller minimal defective $k$-sets, even though the total number of $k$-sets is known to increase with increasing $k$ \cite{clarfeld2019risk}. In most cases tested, both algorithms found 2-sets, 3-sets, and 4-sets in a ratio of roughly 6:3:1, although the proportion of higher order $k$-sets found is notably lower at $T=5$, the case where there are the most defective $k$-sets, and at the smallest initial set size $a_0 = 16$, where the probability of including a higher order $k$-set is lowest. 

\begin{figure}[!ht]
\centering
\includegraphics[width=2.7in]{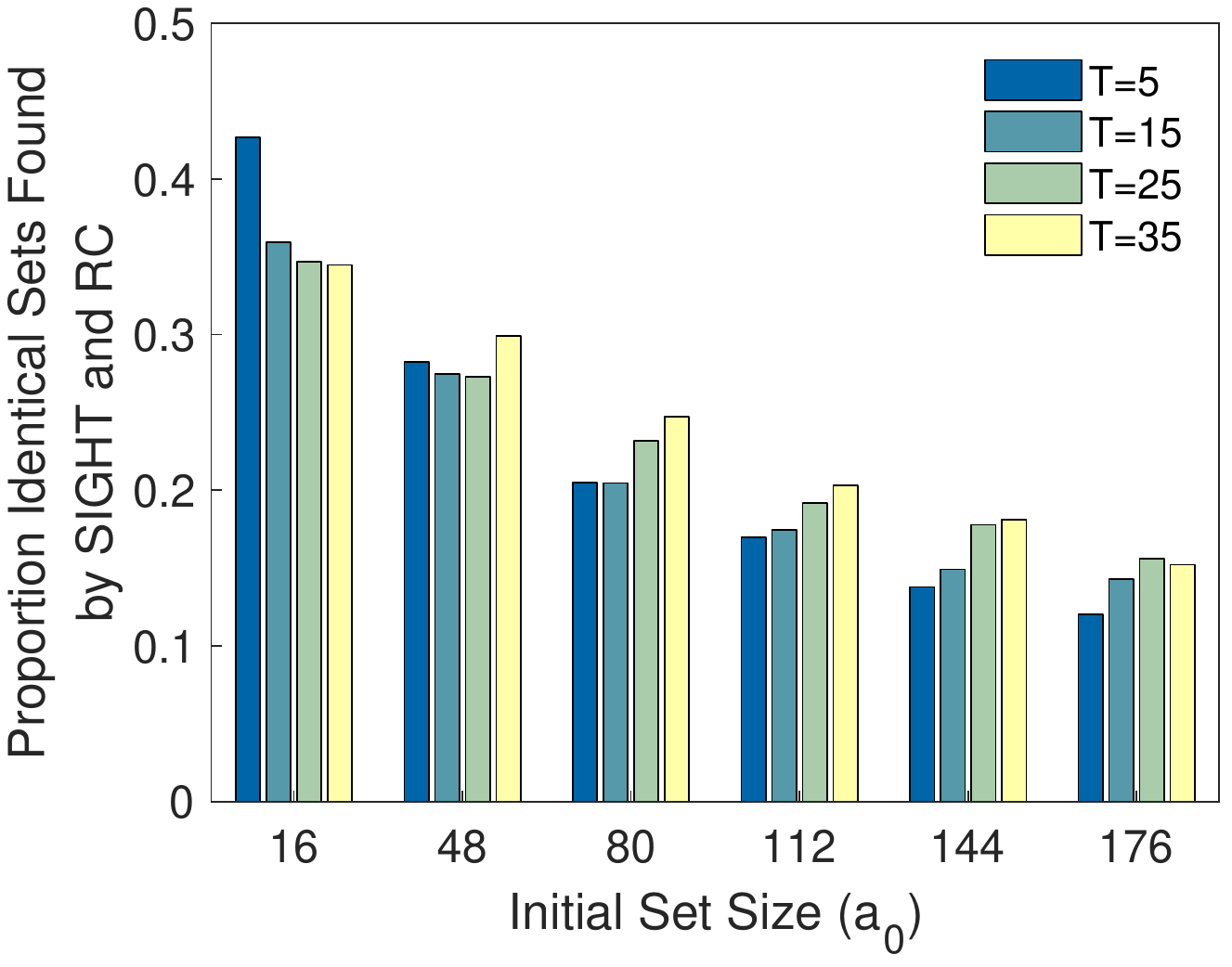}
\caption{The proportion of $k$-sets that were the same in 30,000 paired runs of SIGHT and RC with $k_{max}=4$, for each $a_0 \in \{16,48,80,112,144,176\}$ and each $T \in \{5,15,25,35\}$. }
\label{fig:propsame}
\end{figure}

Despite this similarity in proportions, the two algorithms are not necessarily finding the same minimal defective sets, even though they start with identical random initial sets in each paired run (Fig. \ref{fig:propsame}). This occurs because of the large number of minimal defective $k$-sets present in these test problems and the different set size reduction approaches taken by the two algorithms. Not surprisingly, the proportion of identical minimal defective sets found by SIGHT and RC drops as $a_0$ increases, because the number of minimal defective sets embedded in the initial set increases with increasing $a_0$. Similarly, for $a_0 \ge 48$, the proportion of identical sets found tends to increase with larger $T$, since increasing $T$ reduces the number of minimal defective sets in the system.

\begin{figure*}[!ht]
\centering
\includegraphics[width=4.5 in]{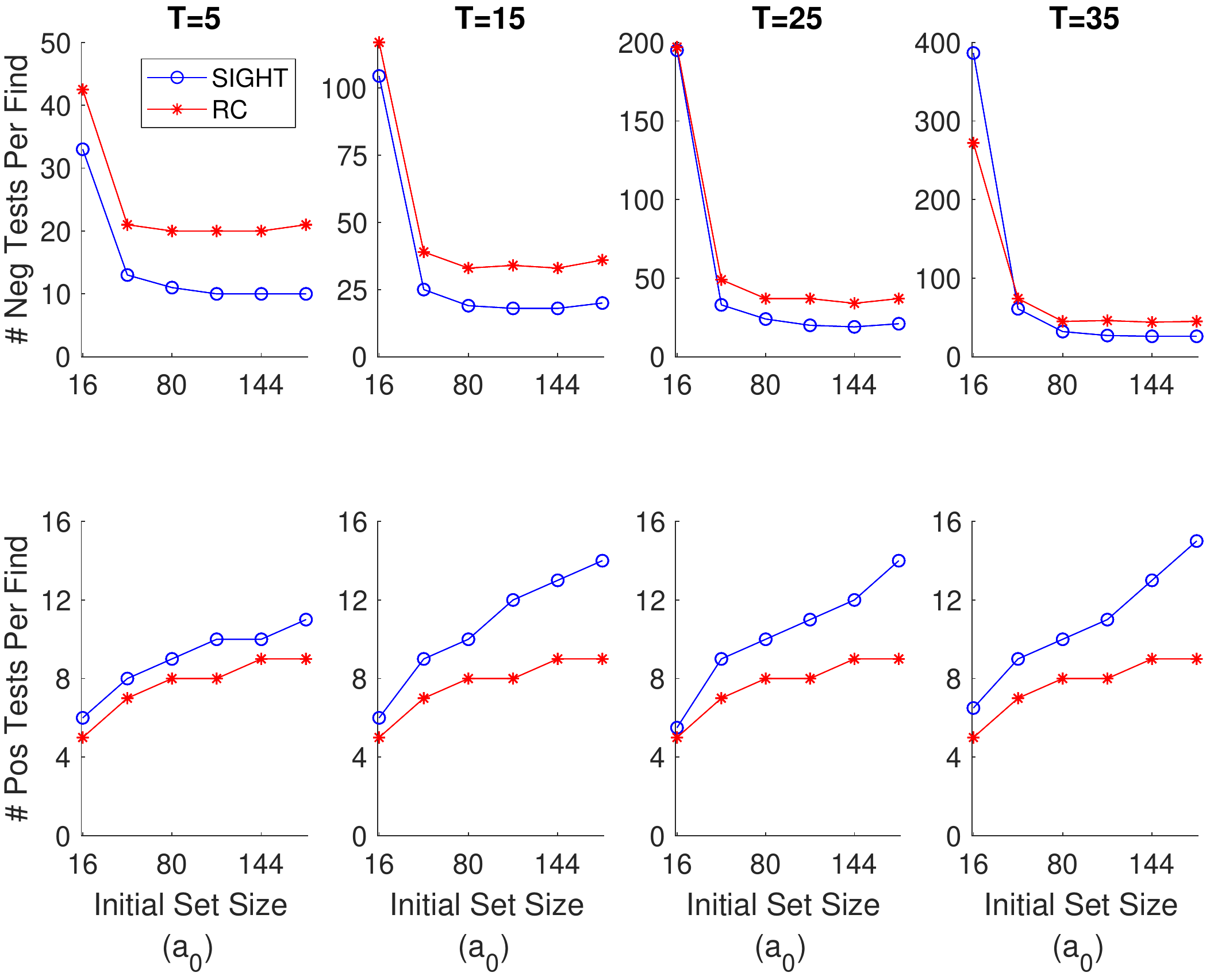}
\caption{Median number of positive and negative tests required per successful find of a minimal defective $k$-set of size $k \in \{2,3,4\}$. Note the differences in scaling of the $y$-axes.}
\label{fig:nPosNeg}
\end{figure*}

\begin{figure*}[!t]
\centering
\includegraphics[width=\textwidth]{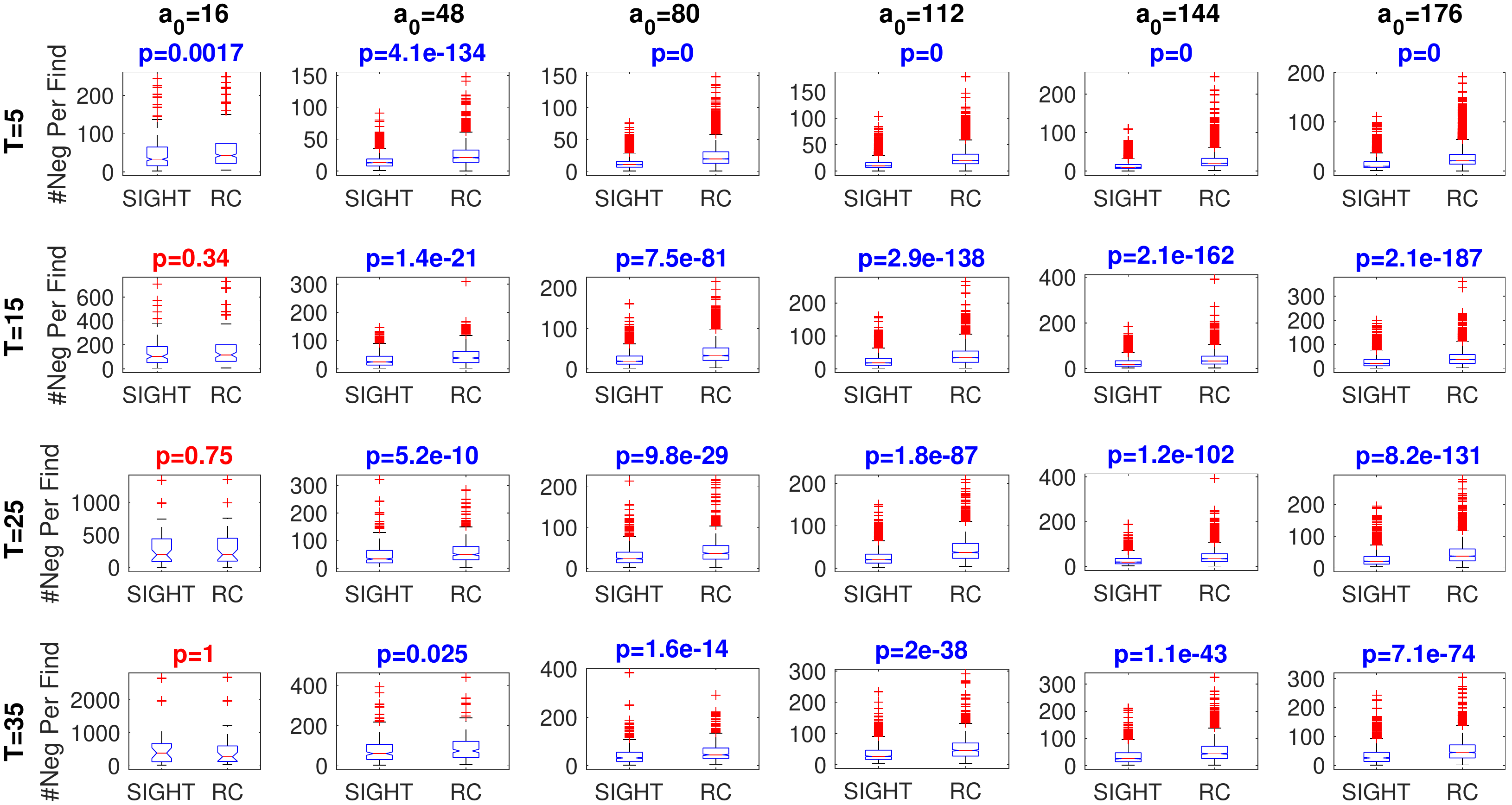}
\caption{Distributions of number of negative tests required per successful find of a minimal defective $k$-set of size $k \in \{2,3,4\}$. Blue $p$-values are statistically significant ($p < 0.05$).  Note the differences in scaling of the $y$-axes. }
\label{fig:NNegPerFind}
\end{figure*}

\begin{figure*}[!t]
\centering
\includegraphics[width=\textwidth]{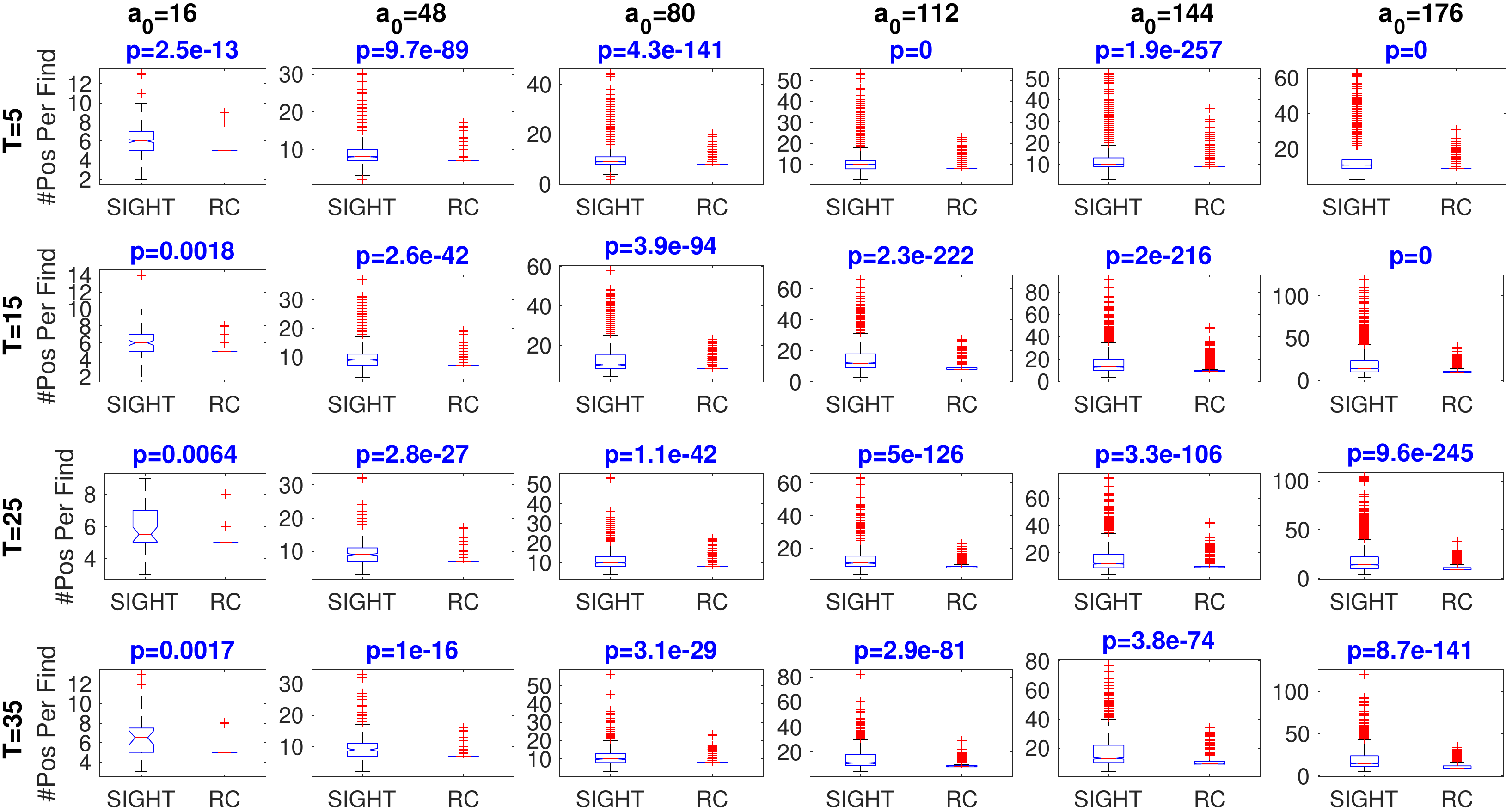}
\caption{Distributions of number of positive tests required per successful find of a minimal defective $k$-set of size $k \in \{2,3,4\}$. Blue $p$-values are statistically significant ($p< 0.05$).  Note the differences in scaling of the $y$-axes.}
\label{fig:NPosPerFind}
\end{figure*}

In nearly all circumstances tested, both algorithms require many more negative tests than positive tests per each successful find (the only exception being for SIGHT with $T=5$ and $a_0>80$, due to the very large number of minimal defective sets present in the initial sets). For both algorithms, the median number of negative tests decreases rapidly as $a_0$ increases from 16 to 48 and then tends to plateau (Fig. \ref{fig:nPosNeg}, top row) while the number of positive tests required increases with increasing $a_0$ (Fig. \ref{fig:nPosNeg}, bottom row). SIGHT nearly always required significantly fewer negative tests than RC and exhibited less variability in the number of negative tests required per find (Fig. \ref{fig:NNegPerFind}). The exception is when $a_0 = 16$ and $T \ge 15$, where the median number of negative tests was not significantly different; this occurs because the initial fail rate was over 99\% and thus dominated the number of negative tests required by both algorithms. However, in all cases tested, SIGHT required significantly more positive tests than RC and exhibited greater variability in the number of positive tests per find (Fig. \ref{fig:NPosPerFind}). 

\begin{figure}[!t]
\centering
\includegraphics[width=2.6in]{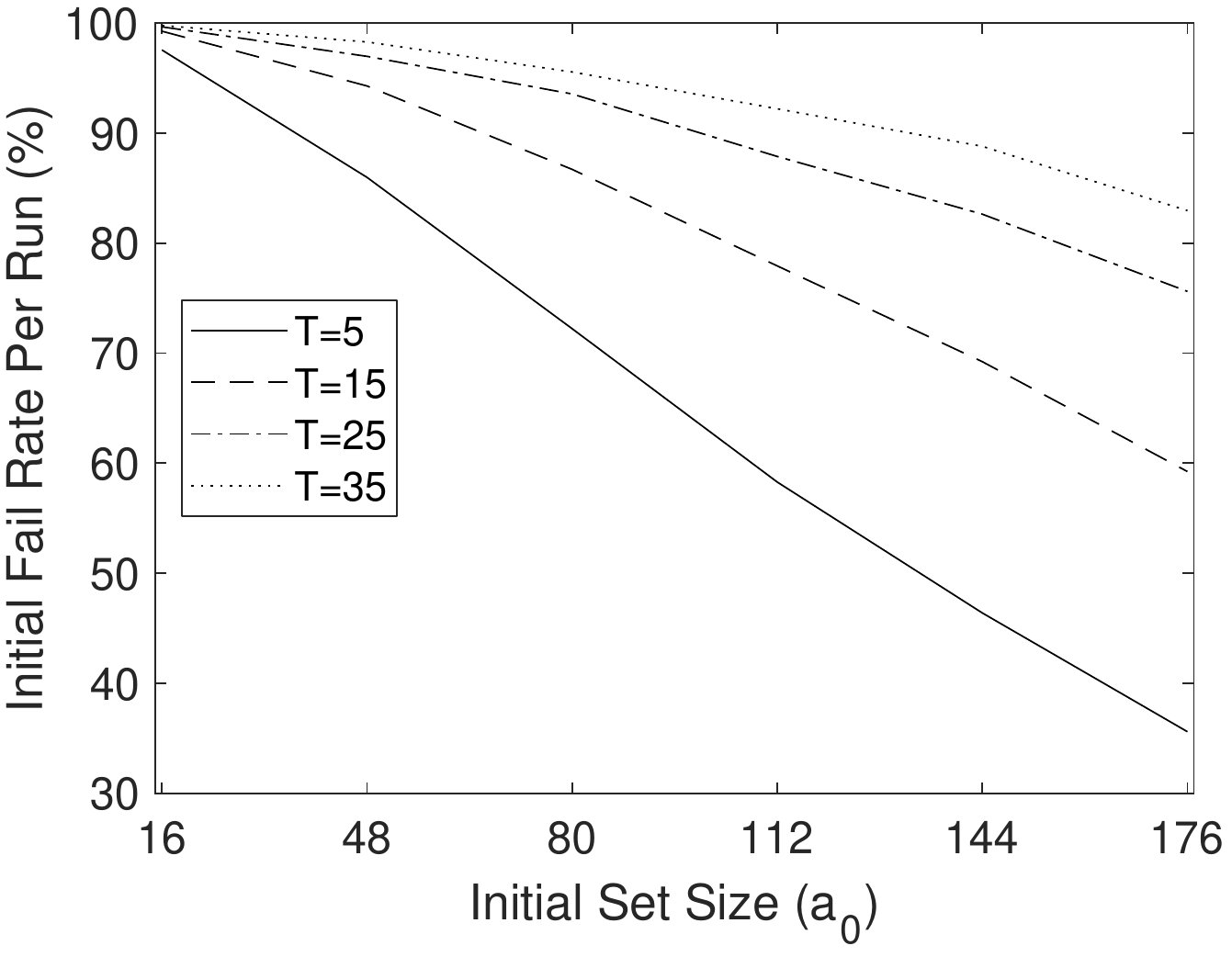}
\caption{The percent of runs that aborted due to the initial random set of size $a_0$ testing as non-defective. Since each paired run of SIGHT and RC started from the same initial random sets, this rate is identical for both algorithms.}
\label{fig:InitFailRate}
\end{figure}

As expected, the initial failure rate ({\em i.e.}, the frequency with which the initial random set of size $a_0$ tests as non-defective) decreases monotonically both as the initial set size increases and as the number of defective sets decreases with increasing $T$ (Fig. \ref{fig:InitFailRate}).  However, in runs where the initial set tests as defective, the failure rate due to subsequent aborts {\em increases} with increasing $a_0$ for both algorithms (Fig. \ref{fig:AbortRate}; see Appendix for a proof of why this occurs). 

Determining the optimal initial set size ({\em i.e.}, where expected runtime is lowest) is non-trivial. As the P:N test cost ratio increases, the optimal initial set size becomes smaller for both algorithms (Fig. \ref{fig:runtimes}, view top to bottom within each column). For a given P:N test cost ratio, the optimal $a_0$ becomes larger as the prevalence of minimal defective $k$-sets shrinks due to increasing $T$ (Fig. \ref{fig:runtimes}, view right to left within each row ). 

For nearly all combinations of $a_0$ and $T$, the median number of total tests was significantly lower ($p < 0.005$) for SIGHT than for RC (Fig. \ref{fig:runtimes}, top row). (The distributions were not significantly different when $T \ge 15$ and $a_0=16$ or when $T=35$ and $a_0 = 48$, because the exceedingly high initial failure rates (Fig. \ref{fig:InitFailRate}) mean that negative tests on the paired initial sets dominate the required number of tests per find for both algorithms.) Thus, if positive and negative tests require the same computational cost, SIGHT is expected to be faster than RC.  However, as the P:N test cost ratio increases, RC is expected to become faster than SIGHT (Fig. \ref{fig:runtimes}, view top to bottom) due to the higher number of positive tests required by SIGHT (Figs. \ref{fig:nPosNeg}, \ref{fig:NPosPerFind}).

\begin{figure}[!t]
\centering
\includegraphics[width=3.5in]{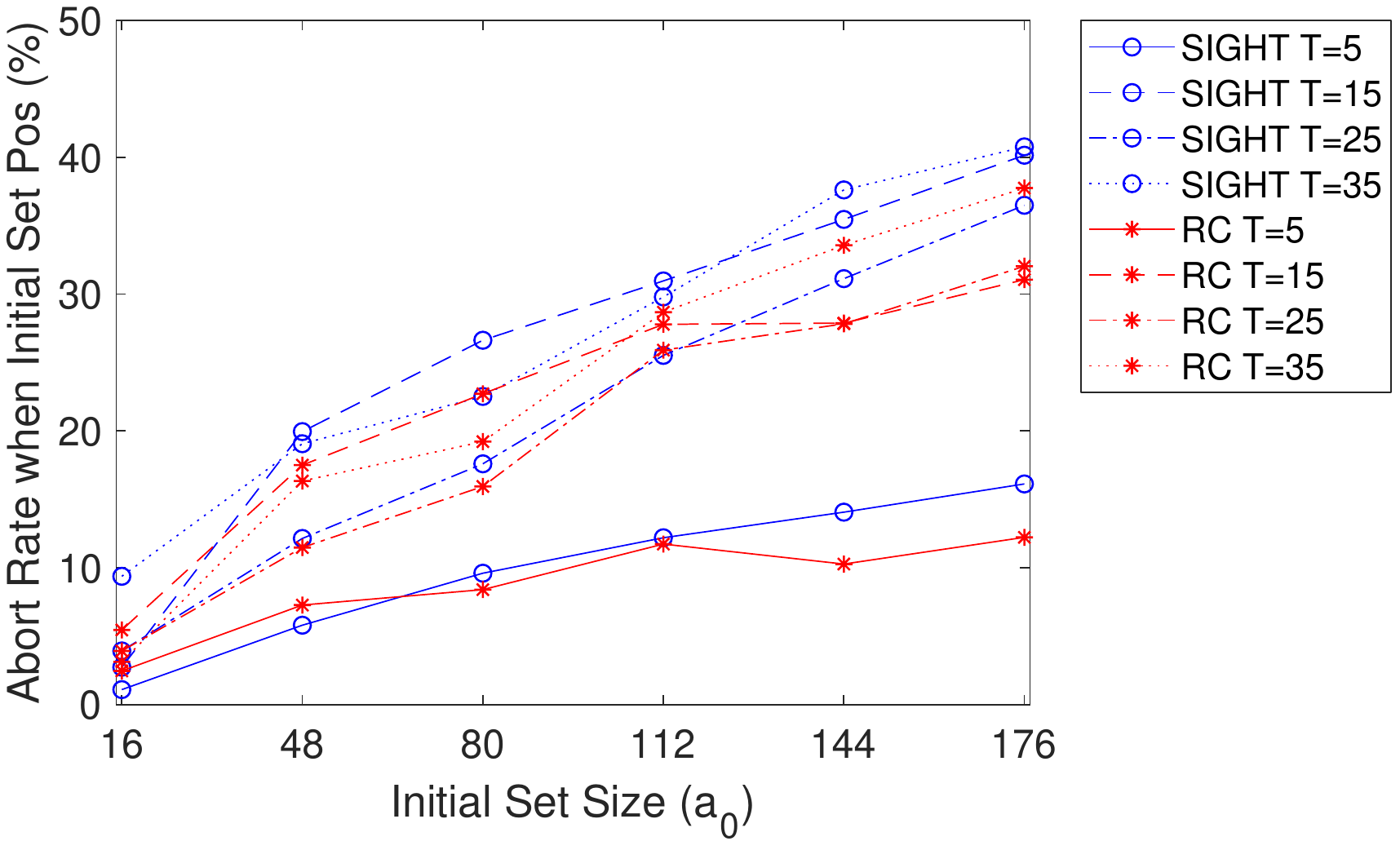}
\caption{The percent of runs in which the initial set of size $a_0$ was found defective but that were subsequently aborted.}
\label{fig:AbortRate}
\end{figure}

\begin{figure*}[!ht]
\centering
\includegraphics[width=5.5 in]{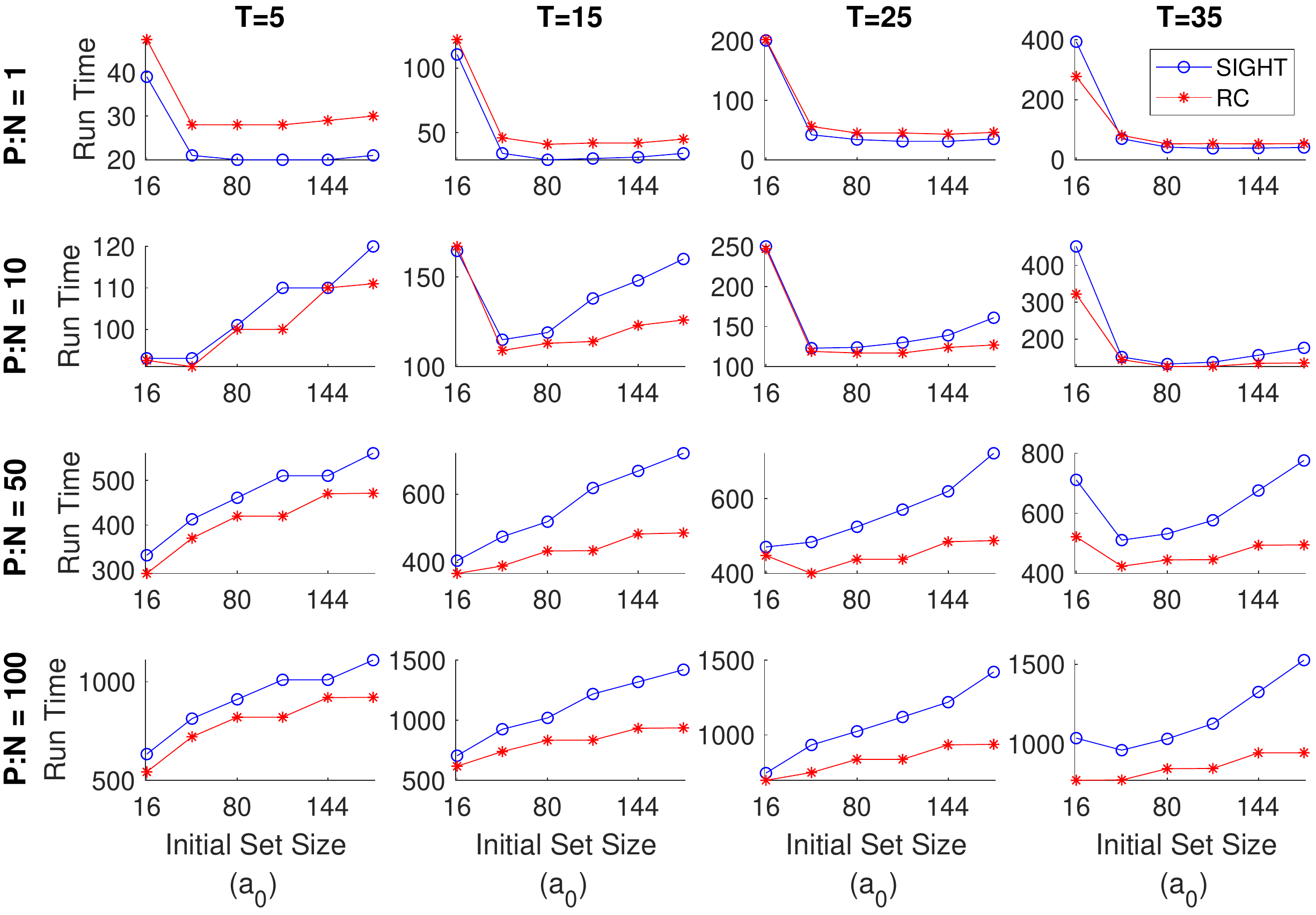}
\caption{Expected computational costs for SIGHT and RC per find at various P:N test cost ratios, for problems generated with the four thresholds $T$. The first row, where P:N $=1$, also represents the median total tests required per find.  Note the differences in scaling of the $y$-axes. }
\label{fig:runtimes}
\end{figure*}

\section{Discussion} \label{sec:discussion}

In this work we compare deterministic (SIGHT) and stochastic (RC) adaptive group-testing algorithms on the problem of sampling minimal defective hyperedges (a.k.a., $k$-sets), for $k \le k_{max}$, in problems where there are intractably large numbers of defective $k$-sets ($k > 1$), false negatives can sometimes occur, and where the relative computational cost of testing defective sets and non-defective sets may differ. 

Both algorithms are similarly effective in finding minimal $2$-sets, $3$-sets, and $4$-sets, with a bias toward finding minimal defective $k$-sets with smaller $k$. While much of this bias is introduced in the random selection of the initial set (because smaller minimal defective $k$-sets are more likely to be included in the initial set), it is amplified during set reduction by different mechanisms in the two algorithms. In RC, with every subset reduction step there is a greater chance of preserving more smaller minimal defective $k$-sets than larger ones, since it is more likely that a larger minimal defective $k$-set will be disrupted during the sub-sampling procedure. Furthermore, the bottom-up search of the set of size $a_{final}$ at the end of RC, will always return one of the smallest contained minimal defective $k$-sets.  In contrast, the binary search step in SIGHT searches for the leftmost element in $S$ that is the rightmost element of a defective $k$-set in $D \concat S$. Thus, the larger a minimal defective set is, the less likely it is to be selected by this procedure. 

As a deterministic algorithm, SIGHT will always identify the same minimal defective $k$-set if starting from the same initial set; this is not the case for RC, where sets are stochastically reduced. In these experiments, the two algorithms found the same minimal defective $k$-sets less than half of the time in paired runs that started from identical initial sets. 

The computational efficiency of both algorithms varies considerably with the initial set size $a_0$, so it is important to try to optimize this control parameter.  It is obvious that, as $a_0$ increases, the chances of aborting a trial due to the initial set being non-defective decreases (assuming false negatives are not predominant), as illustrated in Fig. \ref{fig:InitFailRate}. However, this type of failure is relatively computationally cheap, since it only requires a single test. In contrast, the percent of runs in which the initial set tests as defective but that later abort (due to the minimal defective set being larger than $k_{max}$) increases with increasing $a_0$ (Fig. \ref{fig:AbortRate}; see Appendix for a proof of why this occurs). These latter types of failures can be quite computationally expensive, since they require many tests. The optimal $a_{0}$ finds the right balance between the relative frequencies of these two types of failures.

Our empirical results show that the optimal $a_0$, while similar for both algorithms on the same problem, is problem-specific. For example, the optimal $a_0$ is shown to be smaller for problems that have more defective sets (given the same P:N cost ratio). Similarly, the optimal $a_0$ is shown to be smaller for problems with higher P:N cost ratios (given the same number of defective sets). 

Ultimately, the optimal $a_0$ is determined by the combined effects of (i)  the steady increase in the number of positive tests required per find with increasing $a_0$ (Fig. \ref{fig:nPosNeg}, bottom row), (ii) the sharp drop in the number of negative tests required per find as $a_0$ increases from 16 to 48 (Fig. \ref{fig:nPosNeg}, top row), (iii) the (unknown) number of defective sets present in the system, and (iv) the P:N test cost ratio. The complex interplay of these factors is not possible to predict for real-world problems without simulating them. Thus, we recommend doing a parameter sweep to select an appropriate $a_0$, when applying either SIGHT or RC to a new problem.  

We note that when RC was first presented for the power systems problem, we suggested that $a_0$ be selected to be large enough to achieve a low initial failure rate  \cite{eppstein2012random}. Accordingly, when we first applied RC to the large Western US Test case to assess the risk of cascading failure, $a_0$ was chosen to be 320 in~\cite{clarfeld2018assessing}. However, inspired by the new insights provided by the current study, we later found that reducing $a_0$ from $320$ to $96$ reduced run time per find, when using DCSIMSEP on this system, by nearly half when $k_{max}=4$ (and by even more when $k_{max} < 4$), even though using the smaller $a_0$ increased the initial failure rate from 29\% to 97\%.  Run times would be expected to be even more sensitive to $a_0$ if using a more computationally expensive AC simulator \cite{ju2018modeling}. Assessing risk of cascading failure in a power system requires the identification of {\em many} minimal defective $k$-sets \cite{rezaei2014estimating}, and risk must be reassessed frequently (each time the load level changes) if it is to be mitigated \cite{rezaei2015rapid}. Thus, choosing an appropriate $a_0$ will have a large impact on the computational practicality of using any group sampling approach to assess risk in power systems.

The reliability of tests also has a meaningful impact on the computational efficiency of both algorithms, but for different reasons. In SIGHT, a false negative test may cause an element of a minimal defective $k$-set to initially go undetected in the deterministic binary search portion of SIGHT, such that the defective set eventually found in the search is non-minimal. When this occurs, the element subsequently added to the growing set $D$ is not actually part of a minimal defective set.  This is what necessitates the need for the computationally costly final search of $D$ to ensure that the defective set returned is minimal. 

In contrast, while false negative tests can increase the required number of tests at a given set size for RC, the stochastic nature of subset selection in RC usually enables it to find a defective set despite the presence of false negatives. The bottom-up search of the final set of size $a_{final}$ is always required by RC, whether or not false negatives are present during the subset reduction steps.

The time complexity of successful runs of SIGHT and RC is $O(log N)$ in the number of tests required, for both algorithms. However, this does not mean that their average performances per find are the same. Overall, SIGHT requires fewer total tests than RC but more tests of defective sets, under nearly all circumstances tested. Thus, which algorithm is faster depends on the relative costs of testing defective \textit{vs.} non-defective sets.

In the electric power system application that initially inspired this study, testing defective sets is generally much slower than testing non-defective sets, due to the high computational cost of simulating cascading failures, as shown using DCSIMSEP in Fig. \ref{fig:pos_vs_neg}.   If using a more sophisticated AC simulator, this discrepancy will be even higher, since AC simulators are much slower \cite{ju2018modeling}. Thus, we expect that RC will be faster than SIGHT for realistic problems in this power systems application, regardless of the particular power systems simulator used.  

We note that the important question of whether existing DC or AC simulators are more appropriate for simulating cascading failures in power grids is outside the scope of, and indeed not even relevant to, this work. Our results are independent of what the minimal defective $k$-sets generated in these test problems represent. 

Not all potential applications of these algorithms have higher costs for testing defective sets. For example, group testing has been proposed as a method for performing feature selection for classification tasks \cite{zhou2014parallel}. In this application, tests would determine whether a set of features achieves a specified level of performance on the classification task. ``Defective'' tests, in which a classifier exceeds the performance threshold, may actually be faster than ``non-defective'' tests, since after the threshold is reached classifier training can be aborted.  Thus, it is expected that SIGHT would be (potentially much) faster than RC in the feature selection application. Group testing has also been employed  as a method for discovering synergistic reactions between drugs \cite{remlinger2006statistical,severyn2011parsimonious,hughes2006pooling,borisy2003systematic}, an application in which the cost of tests is constant; hence, since it requires fewer tests, SIGHT would be expected to outperform RC for this application. 

\section{Conclusions} \label{sec:conclusion}
This work compares, for the first time, deterministic (SIGHT) and stochastic (RC) adaptive group-testing methods for sampling minimal defective $k$-sets with variable $k \le k_{max}$, where false negatives may be present, and where the computational costs of testing defective and non-defective sets may differ. We develop a test problem generator that enables us to vary the total number of minimal defective $k$-sets and we use a tunable parameter to control the relative computational costs of testing defective {\em vs.} non-defective sets. 

Like RC, SIGHT is designed to sample from minimal defective sets of variable but bounded $k$, rather than identifying all minimal defective $k$-sets (as in prior works using deterministic group-testing on hypergraphs \cite{chen2007competitive,chen2011revised}).  In addition, the possible presence of false negatives has been largely ignored in the literature on adaptive group testing in graphs/hypergraphs  \cite{li2014pooled,chang2010identification,chang2011pooling}, potentially causing them to fail. The slight modifications necessary to handle these conditions are all that distinguish SIGHT from its group testing ancestors \cite{chen2007competitive,chen2011revised}.

Both RC and SIGHT yielded similar distributions of minimal defective $k$-sets, when searching the same random initial defective sets, and exhibited similar sampling bias toward finding minimal defective sets with lower $k$. The computational efficiency of both algorithms is sensitive to the selection of the initial set size $a_0$, which should be determined empirically since it is shown to be problem dependent. Despite their similarities, the computational properties of these two algorithms are shown to have important differences.

The existing group testing literature has treated all tests as having equal cost, and consequently the standard practice has been to try to develop deterministic algorithms that minimize the number of tests per defective set found \cite{du2000combinatorial}. SIGHT was, in fact, shown to require fewer total tests than RC. However, SIGHT was also shown to require more tests of defective sets than RC. Thus, which method is faster depends on the relative costs of testing defective \textit{vs.} non-defective sets.  

Our results indicate that the stochastic RC group testing algorithm is expected to outperform the deterministic SIGHT group testing algorithm in the power systems application that originally motivated this work, because positive tests are much more computationally costly than negative tests. In other applications, such as feature selection and drug discovery, where the cost of positive tests is less than or equal to the cost of negative tests, SIGHT is expected to out-perform RC.

\appendix  
\label{only_apendix}

\section*{Appendix}
Empirically, the percent of runs that abort partway through a run (after the initial set tested as defective) increases with increasing $a_0$, for both algorithms (see Fig. 8 in the main text).

This occurs because the ratio of minimal defective sets of size $(k+1)$:$k$, in some set $S$, increases as the size of the set increases, for all $k$. To prove this, consider a universal set $V$ of elements and the set $\Omega_k$, which contains all minimal defective $k$-sets, for a given $k$. Consider subset $S \subset V$ where the set sizes $|V|=N$ and $|S| = M$. The expected number of minimal defective $k$-sets in $S$ is $|\Omega_k^S| = {M \choose k}/{N \choose k} \times |\Omega_k|$. Now, consider $T \subset V$ where $|T|=M+c$, for some positive constant $c$. Then, it suffices to show: 

$$\frac{|\Omega_{k+1}^S|}{|\Omega_k^S|} < \frac{|\Omega_{k+1}^T|}{|\Omega_k^T|}$$

\begin{proof}
$$
\frac{
    \nicefrac{{M \choose k+1}}{{N \choose k+1}}|\Omega_{k+1}|
        }
    {
    \nicefrac{{M \choose k}}{{N \choose k}}|\Omega_k|
        } < 
\frac{
    \nicefrac{{M+c \choose k+1}}{{N \choose k+1}}|\Omega_{k+1}|
        }
    {
    \nicefrac{{M+c \choose k}}{{N \choose k}}|\Omega_k|
        }
$$\\
$$
{M \choose k+1}{M+c \choose k} < {M \choose k}{M+c \choose k+1} 
$$\\
$$
\frac{M-k}{k+1}{M \choose k}{M+c \choose k} < \frac{M-k+c}{k+1}{M \choose k}{M+c \choose k}
$$\\
$$M-k < M-k+c$$
\end{proof}

Consequently, the number of expected minimal defective sets in $S$ will increase more slowly for $k \leq k_{max}$ (as in a successful run) than for $k > k_{max}$ (as in a run that aborts because the found defective $k$-set is too large). The net effect is that, even though both RC and SIGHT are biased towards finding smaller $k$-sets, the rapid increase in the number of minimal defective sets that are too large, with increasing $a_0$, results in an increase in the proportion of runs that are aborted part way through a run in which the initial set was defective, for both SIGHT and RC. 

\section*{Acknowledgment}

The authors thank Jeffrey S. Buzas and Damin Zhu for helpful discussions regarding group testing algorithms, which ultimately inspired this work. We also thank Paul D.H. Hines for generously sharing his knowledge of power systems and inspiring us to work on this problem, providing the code for DCSIMSEP, and for helpful suggestions on the manuscript. This work was supported in part by NSF Award Nos.~CNS-1735513, ECCS-1254549, and DGE-1144388. The computational resources provided by the University of Vermont Advanced Computing Core (VACC) are gratefully acknowledged.

\bibliographystyle{unsrt}  

\end{document}